\documentclass[a4paper, 12pt]{article}
\usepackage[utf8]{inputenc}

\usepackage{StyleArticle}      

\title{Add-On Regimes and Their Relevance for Quantifying the Effects of Opioid-Sparing Treatments}

\author{
\small{
C. Stoltenberg\textsuperscript{1,*},
M. Janvin\textsuperscript{1},
M. J. Stensrud\textsuperscript{2},
L. A. Rosseland\textsuperscript{3},
and J. M. Gran\textsuperscript{1}
} \\
\footnotesize{
\textsuperscript{1} Department of Biostatistics, University of Oslo, Norway} \\
\footnotesize{ \textsuperscript{2} Institute of Mathematics, Ecole Polytechnique Fédérale de Lausanne, Switzerland} \\
\footnotesize{\textsuperscript{3} Institute of Clinical Medicine, University of Oslo, Norway} \\
\footnotesize{\textsuperscript{*}\texttt{a.c.stoltenberg@medisin.uio.no}}
}

\date{\small{\today}}

\begin{document}
\doublespacing  

\maketitle
\thispagestyle{empty}
 
\begin{quote}
\centerline{\textbf{Abstract}} 
Medical researchers and practitioners want to know if supplementing opioid treatments with other analgesics, such as nonsteroidal anti-inflammatory drugs (NSAIDs), can reduce opioid consumption. However, quantifying opioid-sparing effects is challenging; even coming up with a policy-relevant estimand requires care. We propose defining these effects in terms of add-on regimes. An add-on regime assigns NSAIDs over time based on the opioid and NSAID treatments a patient would naturally take without any intervention. The regime uses the physician’s decision to administer opioids as a clinically meaningful, and practically feasible, indication for NSAID administration. In contrast, static regimes assign NSAIDs at predefined time points, regardless of clinical context. When opioids are not administered, the add-on regime requires no intervention, thereby preserving the natural level of NSAIDs. This differs from conventional dynamic regimes, which define treatment decisions at every time point during the treatment period. We identify the effect of add-on regimes under assumptions that are easier to assess than those used in existing methods. Finally, we apply the methods to estimate opioid-sparing effects of NSAIDs in a cohort of Norwegian trauma patients using national registry data.
\end{quote}


\pagenumbering{arabic} 

\section{Introduction}
\label{sec:introduction}

\noindent Opioids are effective treatments for managing short-term acute pain \parencite{gaskell2009single, busse2018opioids}. However, their use carries substantial risks, including opioid use disorder, overdose, respiratory depression, falls, and a range of gastrointestinal, endocrine, and immune-related side effects \parencite{vallejo2004opioid, moore2005prevalence, dunn2010opioid, zedler2014risk, bell2018preoperative, rose2018potentially, santosa2020higher, wilson2021preoperative}. Reducing reliance on opioids while maintaining adequate pain management is therefore an important clinical priority.\\
\indent A promising approach is to combine opioids with other analgesics, potentially allowing lower doses of each drug and improving tolerability. Several medications—including nonsteroidal anti-inflammatory drugs (NSAIDs), acetaminophen, clonidine, dexmedetomidine, tizanidine, and ketamine—are under investigation for their opioid-sparing potential \parencite{kumar2017review}. However, quantifying the opioid-sparing effects of such regimes is nontrivial. \\
\indent In this work, we define this effect within a causal (counterfactual outcomes) framework, as an effect of \emph{add-on regimes}, and develop general methodology for causal inference under such regimes. Specifically, we propose to emulate or conduct a randomized target trial \parencite{hernan2016using} comparing two regimes over a defined treatment period: (1) administering the candidate opioid-sparing medication whenever opioids are administered, versus (2) withholding the medication whenever opioids are administered. At times when opioids are not administered, the candidate medication is administered as naturally intended in the absence of intervention. We refer to these regimes as add-on regimes because the candidate medication is added—or not added—whenever opioids are administered. \\
\indent The presented theory is general in that it applies to any setting where the goal is to quantify the effect of implementing one intervention whenever another event of interest occurs, on some outcome. For example, medical researchers are interested in whether prescribing probiotics whenever antibiotics are given reduces antibiotic-associated diarrhea \parencite{hempel2012probiotics}, whether administering anti-nausea medication alongside chemotherapy improves patient comfort \parencite{inrhaoun2012treatment}, and whether offering cognitive behavioral therapy with antidepressants enhances mental health outcomes \parencite{dimidjian2006randomized}. \\
\indent The add-on regime is a time-varying dynamic regime dependent on natural treatment values \parencite{robins2004effects, haneuse2013estimation, richardson2013single, young2014identification, diaz2023nonparametric}. It is dynamic because the decision to administer or withhold NSAIDs at a given time point depends on opioid administration. It depends on natural treatment values because, at any given time point when opioids are not administered, it preserves the NSAID treatment the patient would naturally receive in the absence of intervention at that time point. \\
\indent For our motivating example on opioid-sparing treatments, we argue that the add-on regime represents a class of treatment strategies that are clinically relevant, ethically appropriate, and practically feasible, closely reflecting real-world medical practice in pain management. Static regimes assign NSAIDs at fixed, predefined time points, regardless of whether their use is clinically indicated or feasible in practice. In contrast, the add-on regime uses the physician’s decision to administer opioids—assumed to indicate a clinical need for pain relief—as a meaningful and practical trigger for NSAID administration. \\
\indent Conventional dynamic regimes define treatment decisions, potentially based on the covariate and outcome history, at each time point during the treatment period. For example, one might compare (1) administering NSAIDs if and only if opioids are administered, versus (2) never administering NSAIDs. The add-on regime, on the other hand, preserves the patient’s natural NSAID level at times when opioids are not administered, instead of imposing a rule at every time point. This distinction is clinically important because presumed opioid-sparing medications like NSAIDs are often used for reasons unrelated to opioid treatment. For instance, NSAIDs may be needed to manage fever, inflammation, or other conditions—uses that would be inappropriately blocked under the dynamic regimes described above but remain preserved under the add-on regime. \\
\indent We prove two identification results consistent with prior, more general work \parencite[p.~67]{richardson2013single}, but under a distinct set of assumptions. Specifically, the assumptions of no unmeasured confounding required here are implied by, and involve fewer variables than, those in \textcite[p.~67]{richardson2013single}. Because the new condition involves fewer variables, we argue that it is easier to assess. We also extend the results to handle censoring and competing events.\\ 
\indent The remainder of this article is organized as follows. In \autoref{sec:observeddata}, we describe the observed data structure. In \autoref{sec:regimes}, we define add-on effects as effects of add-on regimes and explain how these effects can be used to quantify opioid-sparing effects in our motivating example. In \autoref{sec:identification}, we identify the add-on effect under sufficient conditions. In \autoref{sec:censor}, we extend the presented theory to handle censoring and competing events. In \autoref{sec:estimation}, we briefly discuss relevant estimation methods for the proposed estimators. Finally, in \autoref{sec:application}, we apply the presented theory to study the opioid-sparing effect of NSAIDs in trauma patients initially treated with opioids.

\section{Observed Data}
\label{sec:observeddata}

\begin{figure}[h]
    \centering
        \begin{tikzpicture}[every path/.style={>=latex}, scale=0.7, transform shape]
            
            \node[draw, shape = ellipse] (L_0) at (0, 6)  {$\boldsymbol{L}_0$};
            \node[draw, shape = ellipse] (L_1) at (4, 6)  {$\boldsymbol{L}_1$};

            \node[draw, shape = ellipse] (Y_0) at (0, 3)  {$Y_0$};
            \node[draw, shape = ellipse] (Y_1) at (4, 3)  {$Y_1$};
            \node[draw, shape = ellipse] (Y_2) at (8, 3)  {$Y_2$};
            
            \node[draw, shape = ellipse] (A_0) at (0, 0)  {$A_0$};
            \node[draw, shape = ellipse] (A_1) at (4, 0)  {$A_1$};

          \draw[->] (L_0) edge (Y_0);
          \draw[->] (L_0) edge [bend right = 20] (A_0);
          \draw[->] (L_0) edge (L_1);
          \draw[->] (L_0) edge (A_1);
          \draw[->] (L_0) edge (Y_1);
          \draw[->] (L_0) edge (Y_2);
          \draw[->] (L_1) edge (Y_1);
          \draw[->] (L_1) edge [bend right = 20] (A_1);
          \draw[->] (L_1) edge (Y_2);

          \draw[->] (Y_0) edge (A_0);
          \draw[->] (Y_0) edge (L_1);
          \draw[->] (Y_0) edge (A_1);
          \draw[->] (Y_0) edge (Y_1);
          \draw[->] (Y_0) [bend left = 15] edge (Y_2);
          \draw[->] (Y_1) edge (A_1);
          \draw[->] (Y_1) edge (Y_2);

          \draw[->] (A_0) edge (L_1);
          \draw[->] (A_0) edge (Y_2);
          \draw[->] (A_0) edge (Y_1);
          \draw[->] (A_0) edge (A_1);

          \draw[->] (A_1) edge (Y_2);
          
        \end{tikzpicture}
\caption{Causal DAG $\mathcal{G}$ illustrating a data generating mechanism for the observed variables.}
\label{fig:DAG}
\end{figure}
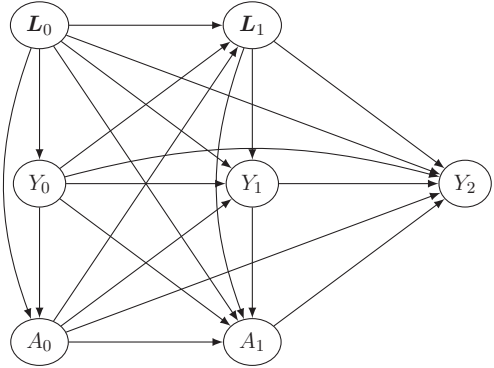

\noindent Let $k \in \{0, \ldots, \kappa, \ldots, K\}$ represent discrete time points (e.g., days) of follow-up. A given treatment regime $g$ is implemented during the treatment period $\{0, 1, \ldots, \kappa\}$ occurring before the end of follow-up $K$. Consider the random variables
\begin{align}
\label{framework:variables}
    (\boldsymbol{L}_0, Y_0, A_0, \boldsymbol{L}_1, Y_1, A_1, \ldots, \boldsymbol{L}_K, Y_K, A_K),
\end{align}
where $\boldsymbol{L}_k \in \mathcal{L}$ is a vector of covariates at time $k$, $Y_k \in \mathcal{Y}$ is the opioid dose at time $k$, and $A_k \in \{0,1\}$ indicate NSAID treatment at time $k$ for every $k \leq K$. The variables are topologically ordered according to (\ref{framework:variables}), defining the linear ordering of the variables as nodes in a graph. \\
\indent To simplify the presentation, we assume no competing events and no censoring until \autoref{sec:censor}. Assume that we have access to $n$ realizations of independent replicates of (\ref{framework:variables}), corresponding to study patients randomly sampled from a target population of interest. A causal directed acyclic graph (DAG) illustrating a data-generating mechanism for the observed variables is presented in \autoref{fig:DAG}. \\
\indent In our motivating example on opioid-sparing treatments, the variables $Y_0, \ldots, Y_{\kappa}$ play a dual role: they serve both as outcome variables (since we are interested in the effect on opioid dose) and as key time-varying covariates (since we are interested in the effect of adding NSAIDs whenever opioids are administered). In other applications, these roles may be represented by distinct variables. For example, when evaluating whether prescribing probiotics alongside antibiotics reduces antibiotic-associated diarrhea, $Y_k$ may denote the outcome of interest, diarrhea at time $k$, while a separate time-varying variable $B_k$ indicates antibiotic use at time $k$. \\
\indent The precise formulation of the research question we pose—and seek to answer—depends on the data at hand. In the application in \autoref{sec:application}, the data on NSAIDs and opioids are obtained from the Norwegian Prescription Database (NorPD), which records all prescription drugs dispensed in Norway for the study population during follow-up. In this setting, the time points correspond to the timing of pharmacy dispensations, so that $Y_k$ represents the opioid dose dispensed at time $k$, and $A_k$ indicates NSAID dispensation at time $k$. Accordingly, the research question implicitly being asked is: “What is the effect of dispensing NSAIDs whenever opioids are dispensed on total opioid dispensing?” In other datasets, the time points might, for example, reflect when the medication was actually used. The implicit research question then becomes: “What is the effect of using NSAIDs whenever opioids are used on total opioid consumption?”. 
\indent We use overbars and underbars to denote the history and future of a sequence of random variables, respectively. Given a sequence $(X_k)_{k \leq K}$ of random variables, denote by $\bar{X}_{k} = (X_{0}, \ldots, X_{k})$ the \textit{history} of $X$ from time $0$ to time $k$, and by $\underline{X}_{k} = (X_{k}, \ldots, X_{K})$ the \textit{future} of $X$ from time $k$ to time $K$. \\
\indent Finally, we introduce the following notation for probability densities. Let $X \in \mathcal{X}$ and $Z \in \mathcal{Z}$ be random variables, potentially multivariate, and let $x \in \mathcal{X}$ and $z \in \mathcal{Z}$. The marginal density of $X$, evaluated at $x$, is denoted by $f_X(x)$, the joint density of $X$ and $Z$, evaluated at $(x,z)$, is denoted by $f_{X,Z}(x,z)$, and the conditional density of $X$ given $Z$, evaluated at $x$ given $Z = z$, is denoted by $f_{X|Z}(x|z)$.

\section{Defining the Add-On Effect}
\label{sec:regimes}

\noindent In this section, we define the add-on effect as an effect of add-on regimes. In our motivating example on opioid-sparing treatment, we use the add-on effect to define the opioid-sparing effect of supplementing opioid treatments with NSAIDs. \\
\indent Following \textcite{robins1986new}, \textcite{robins2004effects}, and \textcite{richardson2013single}, we define a general treatment regime $g$ by a sequence of random variables $(A_k^{g+})_{k \leq \kappa}$, termed the \textit{regime-assigned treatment variables}, where $A_k^{g+} \in \{0,1\}$ indicates NSAID treatment under regime $g$ at each time point $k$ during the treatment period $\{0, 1, \ldots, \kappa\}$. Consequently, under a given regime $g$, we distinguish between two factual variables related to NSAID at each time point: the \textit{observed} NSAID indicator $A_k$, representing NSAID treatment in the absence of an intervention, and the regime-assigned NSAID indicator $A_k^{g+}$, representing the NSAID treatment specified by regime $g$. The observed NSAID levels are assumed to be realizations of $\bar{A}_K$, while $\bar{A}^{g+}_{\kappa}$ is defined by the researcher according to the treatment regime of interest. \\
\indent For example, a regime $g$ is said to be \textit{static} if it is fully specified by a fixed sequence of treatment decisions, i.e., $A_k^{g+} = a_k$ for all $k \leq \kappa$, where $\bar{a}_{\kappa} \in \{0,1\}^{\kappa+1}$ defines the sequence. In this case, we refer to $g$, or simply $\bar{a}_{\kappa}$, as a static regime. \\
\indent Let superscripts denote counterfactual variables. In particular, $Y_k^{\bar{a}_{k-1}}$ is the counterfactual opioid dose at time $k$ when NSAID treatment is fixed by the static regime $\bar{a}_{k-1}$ for every $\bar{a}_{k-1} \in \{0,1\}^k$ and $k \geq 1$. More generally, we let
\begin{align}
\label{def:counter}
    Y_k^{g} := \sum_{ a_0 = 0 }^1 \ldots \sum_{ a_{\min \{k-1, \kappa\}} = 0 }^1 1\{ \bar{A}^{g+}_{\min \{k-1, \kappa\}} = \bar{a}_{\min \{k-1, \kappa\}} \} Y_k^{\bar{a}_{\min \{k-1, \kappa\}}} \quad \forall \; k \geq 1, 
\end{align}
be the counterfactual opioid dose at time $k$ under an arbitrary regime $g$, where NSAID treatment is specified by $\bar{A}_{\kappa}^{g+}$ for every $k \geq 1$. Definition (\ref{def:counter}) explicate that the counterfactual opioid dose under a general regime $g$ is a function of the regime-assigned treatment variables and counterfactual variables under \emph{static} regimes. This is important in our arguments because it clarifies why the identifiability conditions in \autoref{sec:identification} rely on counterfactual variables under static regimes rather than add-on regimes. For a formal treatment of all relevant counterfactual variables, see Appendix A. \\
\indent Following \textcite{richardson2013single}, we refer to $A_k^g$ as the \emph{natural treatment value} under regime $g$ at time $k$ for every $k \geq 1$. Specifically, the natural treatment $A_k^g$ at time $k$ is the treatment that would be administered at time $k$ in the absence of intervention at that time, had the regime-assigned treatment $\bar{A}_{k-1}^{g+}$ through time $\min\{k-1, \kappa\}$ been assigned according to regime $g$. \\
\indent We are now ready to introduce \textit{add-on regimes} through a specific definition of the regime-assigned treatment variables $\bar{A}_{\kappa}^{g+}$. 
\begin{definition}[Add-on regime]
\label{regime:add-on}
Let $g$ be a treatment regime specified by $(A^{g+}_k)_{k \leq \kappa}$. Let $j \in \{0,1\}$ and $g: \mathcal{Y} \times \{0,1\} \to \{0,1\}$. If $A_0^{g+} = g(Y_0, A_0)$ and 
\begin{align}
\label{regime:add-on:formula}
    A_k^{g+} = g(Y^g_k, A^g_k) \quad \text{with} \quad g(y, a) = 
    \begin{cases}  
    j \quad \, &\textup{if } \; y > 0, \\
    a \quad  \, &\textup{if } \; y = 0,
    \end{cases} \quad \forall \; k \leq \kappa, 
\end{align}
we say that $g$ is an add-on-$j$ regime.
\end{definition}
\indent For the remainder of this article, we let $g_1$ denote the add-on-$1$ regime and $g_0$ denote the add-on-$0$ regime. \\
\indent The add-on regime is a dynamic regime dependent on natural treatment values \parencite{robins2004effects,  haneuse2013estimation, richardson2013single, young2014identification,  diaz2023nonparametric}. It is dynamic because the decision to administer or withhold NSAIDs at a given time point depends on the counterfactual opioid dose. It depends on natural treatment values because it preserves the naturally intended NSAID level when opioids are not administered.\\
\indent To illustrate how an add-on regime is implemented in practice, consider a doctor determining pain medication for a patient by following the add-on-$1$ regime over a two-day treatment period $\{0, 1\}$. On day 0, the doctor first elicits the natural opioid dose $Y_0$ and the natural NSAID indicator $A_0$. For example, the doctor could answer the following questions: What opioid dose would you prescribe to this patient if deciding independently? And would you prescribe NSAIDs? The responses to these hypothetical questions define the natural values of $Y_0$ and $A_0$, under the assumption that they reflect the choices the doctor would make in the absence of an intervention. The doctor then prescribes the observed opioid dose $Y_0$ and assigns NSAIDs according to the regime-assigned level $A_0^{g+} = g(Y_0, A_0)$, where the function $g$ is defined by equation~\eqref{regime:add-on:formula} in \autoref{regime:add-on} for $j = 1$.\\
\indent On day 1, the doctor observes the counterfactual natural opioid dose $Y_1^g$ and NSAID indicator $A_1^g$, which represent the natural values that would arise had the NSAID assignment on day $0$ been made according to the add-on-$1$ regime. Based on these values, the doctor prescribes the opioid dose $Y_1^g$ and NSAIDs according to $A_1^{g+} = g(Y_1^g, A_1^g)$, again using the function $g$ defined in equation \eqref{regime:add-on:formula} in \autoref{regime:add-on}. \\
\indent We are now ready to define the \emph{add-on effect}. We say that the treatment has an add-on effect on the outcome at time $k$ if $E\left[ Y_k^{g_1} \right] \neq E\left[ Y_k^{g_0} \right]$ for every $k \geq 1$, indicating that the expected outcome under the add-on-$1$ regime is different from that under the add-on-$0$ regime. The add-on effect is then given by 
\begin{align}
\label{eff:aose:timek}
E\left[ Y_k^{g_1} \right] - E\left[ Y_k^{g_0} \right] \quad \forall \, k \geq 1.
\end{align}
\indent Specifically, for our motivating example on opioid-sparing treatments, we define the \textit{opioid-sparing effect} of supplementing opioid treatments with NSAIDs as
\begin{align}
\label{eff:aose}
\sum_{k=1}^{K} \left( E\left[ Y_k^{g_1} \right] - E\left[ Y_k^{g_0} \right] \right).
\end{align}
\noindent We say that NSAIDs have an opioid-sparing effect if (\ref{eff:aose}) is negative, indicating that the expected total opioid dose over follow-up is lower under the add-on-$1$ regime, where NSAIDs are added to every opioid treatment during the treatment period, than under the add-on-$0$ regime, where NSAIDs are actively withheld during the treatment period. \\
\indent We conclude this section by clarifying why, in our motivating example, the add-on regime is more relevant for policy and clinical practice than conventional static or dynamic regimes. One alternative to our definition of the opioid-sparing effect in \eqref{eff:aose} is the contrast
\begin{align}
\label{eff:static}
\sum_{k=1}^{K} \left( E\left[ Y_k^{\bar{a} = \bar{1}} \right] - E\left[ Y_k^{\bar{a} = \bar{0}} \right] \right),
\end{align}
which compares expected total opioid use under two static strategies: always administering NSAIDs ($\bar{a}\kappa = \bar{1}$) versus never doing so ($\bar{a}\kappa = \bar{0}$). Although a negative value of \eqref{eff:static} would suggest an opioid-sparing effect, the contrast has limited practical relevance. Recommending NSAIDs at every time point, regardless of clinical context, is neither feasible nor advisable. The add-on regime, by contrast, uses opioid administration as a proxy for pain requiring intervention. \\
\indent Another option is to compare a dynamic regime $g$ to the never-NSAID static strategy $\bar{a}_{\kappa} = \bar{0}$:
\begin{align}
\label{eff:dynamic}
\sum_{k=1}^{K} \left( E\left[ Y_k^{g} \right] - E\left[ Y_k^{\bar{a} = \bar{0}} \right] \right),
\end{align}
where $g$ is specified by $A_k^{g+} = 1\{Y_k^g > 0\}$—that is, NSAIDs are administered only when opioids are. While similar to the add-on regime, this regime restricts NSAID use solely to instances of opioid administration, thereby precluding their use for other indications like fever or inflammation—uses that remain allowed under the add-on regime.

\section{Identifying Effects of Add-On Regimes}
\label{sec:identification}

\noindent In this section, we establish sufficient conditions for identifying the add-on effect (\ref{eff:aose:timek}) as functionals of observed data (\ref{framework:variables}).  Identification of these expectations directly implies identification of the opioid-sparing effect (\ref{eff:aose}). Although our goal is to identify expectations of the counterfactual outcome under dynamic add-on regimes, the conditions are formulated with respect to counterfactual variables under static regimes. This is valid because counterfactual variables under add-on regimes are defined in terms of those under static regimes (see \autoref{def:counter} and Appendix A). All proofs are provided in Appendix A.

\subsection{Identification Assumptions}

We assume that the following conventional causal consistency assumption holds. 
\begin{assumption}[Consistency]
\begin{align}
\label{ass:con}
\begin{split}
    \bar{A}_{ k' } = \bar{a}_{ k' } \; \; \Rightarrow \; \;  \boldsymbol{L}_k = \boldsymbol{L}_k^{\bar{a}_{ k' }}, \; Y_k = Y_k^{\bar{a}_{ k'}}, \; A_k = A_k^{\bar{a}_{ k' }} \quad \forall \, \bar{a}_{\kappa} \in \{0,1\}^{\kappa + 1}, \; \forall \, \; k \geq 1,
\end{split}
\end{align} 
where $k' = \min \{k-1, \kappa\}$.
\end{assumption} 
\noindent The consistency assumption states that, for each individual, the observed data is equal to the counterfactual data under the static regime defined by the NSAID treatments the individual actually received. For example, if an individual received NSAIDs at time $0$ with $A_0 = 1$, then their counterfactual opioid dose $Y_1^{1}$ at time $1$ under the static regime $a_0 = 1$ is equal to their observed opioid dose $Y_1$ at time $1$. \\
\indent Violations of consistency can occur if multiple versions of NSAID treatment exist in the data and these versions have different effects on future opioid and NSAID levels \parencite{vanderweele2013causal, vanderweele2022constructed}. For example, if certain subgroups receive NSAID doses alongside opioids that deviate substantially from those given to the general population, and future opioid and NSAID levels are highly sensitive to these dose differences, then the counterfactual opioid doses $Y_k^{\bar{a}}$ and NSAID indicators $A_k^{\bar{a}}$ would not be well-defined. As a result, the counterfactual opioid doses under add-on regimes would also be ill-defined, making the add-on effect (\ref{eff:aose:timek}) undefined. \\
\indent The consistency assumption (\ref{ass:con}) implies that, in the observed data, we will have observations of 
$(\bar{\boldsymbol{L}}_K^{g_j}, \bar{Y}_K^{g_j}, \bar{A}_K^{g_j})$
for all study individuals who actually followed the add-on-$j$ regime $g_j$, since, for them, their observed data $(\bar{\boldsymbol{L}}_K, \bar{Y}_K, \bar{A}_K)$ coincide with their counterfactual data $(\bar{\boldsymbol{L}}_K^{g_j}, \bar{Y}_K^{g_j}, \bar{A}_K^{g_j})$. However, for individuals who did not follow the add-on-$j$ regime $g_j$, data on $(\bar{\boldsymbol{L}}_K^{g_j}, \bar{Y}_K^{g_j}, \bar{A}_K^{g_j})$ will be unavailable. \\
\indent We impose the assumption of no unmeasured confounding through the following sequential exchangeability condition. In Appendix A, we show that this condition is implied by the corresponding assumption used in \textcite[p.~67]{richardson2013single}. Furthermore, we demonstrate that it involves fewer variables than the condition used in \textcite[p.~67]{richardson2013single}, making it easier to assess in applied settings. We first define the following sets
\begin{align}
\label{Wtk}
    &W_{t,k}^{g_j} = (Y_k^{g_j}, Y_{t+1}^{g_j}, \ldots, Y_{ k' }^{g_j}, A_{t+1}^{g_j}, \ldots, A_{ k' }^{g_j} ) \, \cap \, \text{an}_{\mathcal{G}({g_j})}(Y_k^{g_j}), 
\end{align} 
and $W_{j, t,k}^{\bar{a}_{k'}} = \{V^{\bar{a}} \mid V^{g_j} \in W_{t,k}^{g_j} \}$, the subset of vertices in $\mathcal{G}(\bar{a}_{k'})$ that correspond to the vertices in (\ref{Wtk}) for all $j \in \{0,1\}$, $t \leq k'$, $\bar{a}_{k'} \in \{0,1\}^{k' + 1}$, and $k \geq 1$, where $k' = \min \{k-1, \kappa\}$.

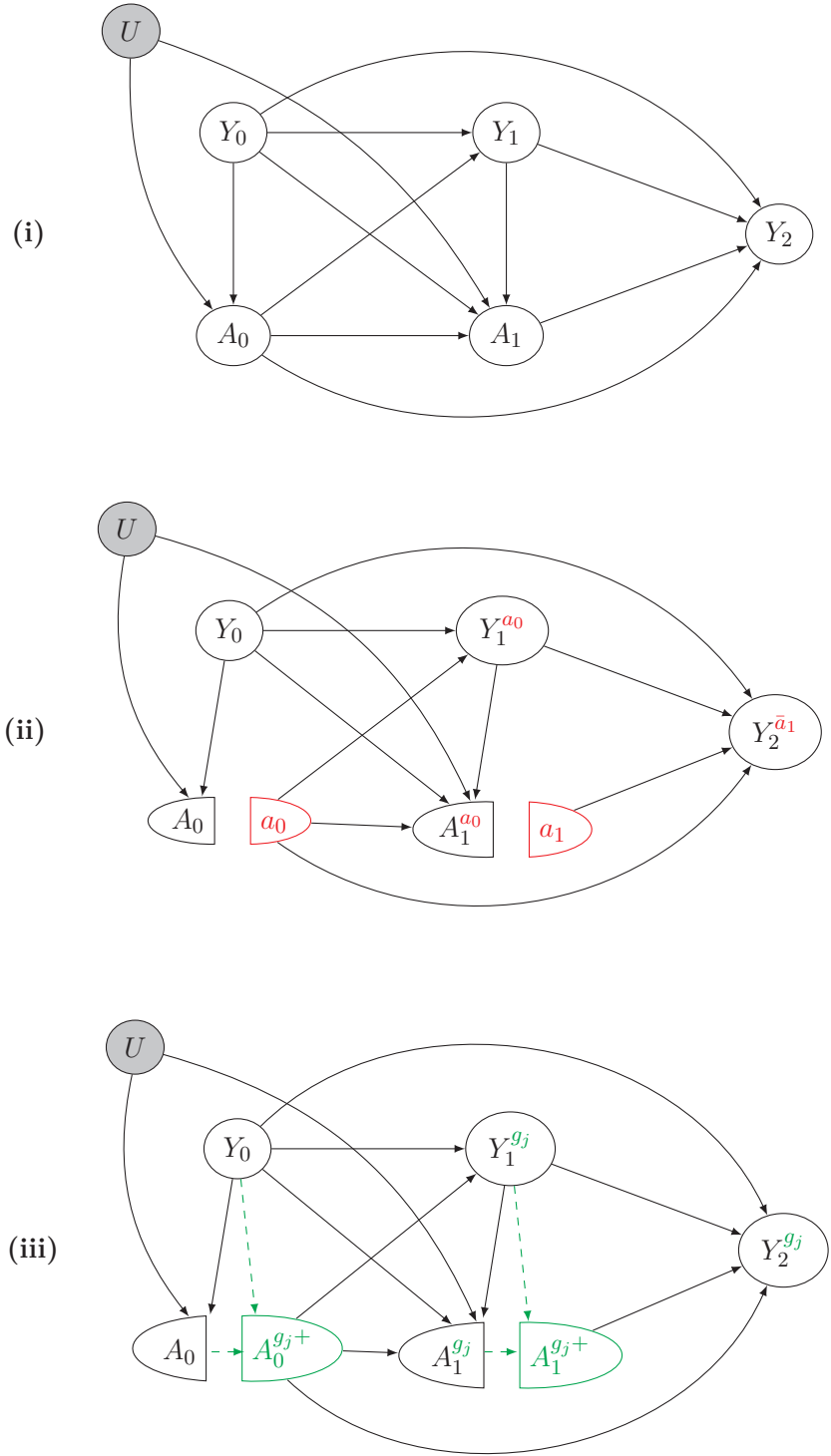
\begin{figure}[p] 
    \centering
    \begin{tikzpicture}[every path/.style={>=latex}, scale=0.9, transform shape]
        \node[] (i) at (-3, 1.5)  {$\textbf{(i)}$};
    
        \node[draw, shape = ellipse, fill=lightgray] (U) at (-1.5, 4.5)  {$U$};

        \node[draw, shape = ellipse] (Y_0) at (0, 3)  {$Y_0$};
        \node[draw, shape = ellipse] (Y_1) at (4, 3)  {$Y_1$};
        \node[draw, shape = ellipse] (Y_2) at (8, 1.5)  {$Y_2$};
            
        \node[draw, shape = ellipse] (A_0) at (0, 0)  {$A_0$};
        \node[draw, shape = ellipse] (A_1) at (4, 0)  {$A_1$};

          \draw[->] (U) edge [bend right = 20] (A_0);
          \draw[->] (U) edge [bend left = 20] (A_1);

          \draw[->] (Y_0) edge (A_0);
          \draw[->] (Y_0) edge (A_1);
          \draw[->] (Y_0) edge (Y_1);
          \draw[->] (Y_0) [bend left = 45] edge (Y_2);
          \draw[->] (Y_1) edge (A_1);
          \draw[->] (Y_1) edge (Y_2);

          \draw[->] (A_0) [bend right = 45] edge (Y_2);
          \draw[->] (A_0) edge (Y_1);
          \draw[->] (A_0) edge (A_1);

          \draw[->] (A_1) edge (Y_2);
    \end{tikzpicture}
    \vspace{1em}

    \begin{tikzpicture}[every path/.style={>=latex}, scale=0.9, transform shape]
        \tikzset{swig vsplit={gap=15pt, line color right=red}}

            \node[] (ii) at (-3, 1.5)  {$\textbf{(ii)}$};
        
            \node[draw, shape = ellipse, fill=lightgray] (U) at (-1.5, 4.5)  {$U$};

            \node[draw, shape = ellipse] (Y_0) at (0, 3)  {$Y_0$};
            \node[draw, shape = ellipse] (Y_1) at (4, 3)  {$Y_1^{\textcolor{red}{a_0}}$};
            \node[draw, shape = ellipse] (Y_2) at (8, 1.5)  {$Y_2^{\textcolor{red}{\bar{a}_1}}$};
            
            \node[name=A_0,below=20mm of Y_0, shape=swig vsplit]{
                    \nodepart{left}{$A_0$}
                    \nodepart{right}{$\textcolor{red}{a_0}$} };
            \node[name=A_1,below=20mm of Y_1, shape=swig vsplit]{
                    \nodepart{left}{$A_1^{\textcolor{red}{a_0}}$}
                    \nodepart{right}{$\textcolor{red}{a_1}$} };

            \draw[->] (U) edge [bend right = 25] (A_0);
            \draw[->] (U) edge [bend left = 25] (A_1);
    
            \draw[->] (Y_0) edge (-0.4, 0.55);
            \draw[->] (Y_0) edge (A_1);
            \draw[->] (Y_0) edge (Y_1);
            \draw[->] (Y_0) [bend left = 45] edge (Y_2);
            \draw[->] (Y_1) edge (3.6, 0.5);
            \draw[->] (Y_1) edge (Y_2);
    
            \draw[->] (A_0) [bend right = 45] edge (Y_2);
            \draw[->] (A_0) edge (Y_1);
            \draw[->] (A_0) edge (A_1);

            \draw[->] (A_1) edge (Y_2);
    \end{tikzpicture}
    \vspace{1em}

    \begin{tikzpicture}[every path/.style={>=latex}, scale=0.9, transform shape]
        \tikzset{swig vsplit={gap=15pt, line color right=green}}
            \node[] (iii) at (-3, 1.5)  {$\textbf{(iii)}$};
        
            \node[draw, shape = ellipse, fill=lightgray] (U) at (-1.5, 4.5)  {$U$};

            \node[draw, shape = ellipse] (Y_0) at (0, 3)  {$Y_0$};
            \node[draw, shape = ellipse] (Y_1) at (4, 3)  {$Y_1^{\textcolor{green}{g_j}}$};
            \node[draw, shape = ellipse] (Y_2) at (8, 1.5)  {$Y_2^{\textcolor{green}{g_j}}$};
            
            \node[name=A_0,below=20mm of Y_0, shape=swig vsplit]{
                    \nodepart{left}{$A_0$}
                    \nodepart{right}{$\textcolor{green}{A_0^{g_j+ }}$} };
            \node[name=A_1,below=20mm of Y_1, shape=swig vsplit]{
                    \nodepart{left}{$A_1^{\textcolor{green}{g_j}}$}
                    \nodepart{right}{$\textcolor{green}{A_1^{g_j +}}$} };

            \draw[->] (U) edge [bend right = 25] (A_0);
            \draw[->] (U) edge [bend left = 25] (A_1);
    
            \draw[->] (Y_0) edge (-0.4, 0.55);
            \draw[dashed, green, ->] (Y_0) edge (A_0);
            \draw[->] (Y_0) edge (A_1);
            \draw[->] (Y_0) edge (Y_1);
            \draw[->] (Y_0) [bend left = 55] edge (Y_2);
            \draw[dashed, green, ->] (Y_1) edge (A_1);
            \draw[->] (Y_1) edge (3.6, 0.5);
            \draw[->] (Y_1) edge (Y_2);
    
            \draw[->] (A_0) [bend right = 55] edge (Y_2);
            \draw[->] (A_0) edge (Y_1);
            \draw[->] (A_0) edge (A_1);
            \draw[dashed, green, ->] (-0.38,0) to (0.1, 0);

            \draw[->] (A_1) edge (Y_2);
            \draw[dashed, green, ->] (3.62,0) to (4.1, 0);
    \end{tikzpicture}
    
    \caption{(i) A causal DAG $\mathcal{G}$ illustrating a data-generating mechanism. (ii) The SWIT $\mathcal{G}(\bar{a}_1)$, for $\bar{a}_1 \in \{0,1\}^2$, representing static interventions on $A_0$ and $A_1$ within the DAG in (i). (iii) The dSWIT $\mathcal{G}(g_j)$, for $j \in \{0,1\}$, illustrating dynamic interventions on $A_0$ and $A_1$ under the add-on-$j$ regime $g_j$ defined in (\ref{regime:add-on}) within the DAG $\mathcal{G}$ in (i).}
    \label{fig:combined-dag}
\end{figure}

\begin{assumption}[Exchangeability]
\label{ass:ex} 
\begin{align}
\label{ass:ex:formula}
    & W_{j, t,k}^{\bar{a}_{k'}} \; \indep \; A^{\bar{a}_{t-1}}_t \mid \bar{\boldsymbol{L}}_t^{\bar{a}_{t-1}}, \bar{Y}_t^{\bar{a}_{t-1}}, \bar{A}_{t-1}^{\bar{a}_{t-2}},
\end{align}
for all $j \in \{0,1\}$, $t \leq k'$, $\bar{a}_{k'} \in \{0,1\}^{k' + 1}$, and $k \geq 1$, where $k' = \min \{k-1, \kappa\}$. 
\end{assumption}
\noindent To verify exchangeability in (\ref{ass:ex:formula}), we proceed in two steps using two distinct graphs. For each $k \geq 1$ with $k' = \min \{k-1, \kappa\}$ and each $j \in \{0,1\}$, we perform the following: 
\begin{enumerate}[noitemsep]
    \item First, in the dynamic single world intervention template (dSWIT) $\mathcal{G}(g_j)$, we identify the set $W^{g_j}_{t,k}$ of ancestors of $Y_k^{g_j}$ that are also elements of $(Y^{g_j}_{t+1}, A^{g_j}_{t+1}, \ldots, Y^{g_j}_{k'}, A^{g_j}_{k'})$, for all $t \leq k'$. \\
    \item Second, we construct the corresponding set $W_{t,k}^{\bar{a}_{k'}}$ in the single world intervention template (SWIT) $\mathcal{G}(\bar{a}_{k'})$, and assess whether $A_t^{\bar{a}_{t-1}}$ is d-separated from $W_{t,k}^{\bar{a}_{k'}}$ given $\bar{\boldsymbol{L}}_t^{\bar{a}_{t-1}}$, $\bar{Y}_t^{\bar{a}_{t-1}}$, and $\bar{A}_{t-1}^{\bar{a}_{t-2}}$, for every $t \leq k'$ and every $\bar{a}_{k'} \in \{0,1\}^{k'+1}$.
\end{enumerate}
\noindent \autoref{ass:ex} can be violated if there exist unmeasured common causes of NSAID treatment at different times. As an example, consider the DAG $\mathcal{G}$ in \autoref{fig:combined-dag}(i) with an unmeasured common cause $U$ of NSAID treatment $A_0$ and $A_1$. In the doctor–patient setting, this can occur if the physician’s decisions to prescribe NSAIDs on day $0$ and day $1$ are influenced by an unmeasured patient characteristic $U$. From the corresponding dSWIT in \autoref{fig:combined-dag}(iii), we find that $W^{g_j}_{0,2} = \{ Y_1^{g_j}, A_1^{g_j}, Y_2^{g_j} \}$ and $W^{g_j}_{1,2} = \{ Y_2^{g_j} \}$. Examining the corresponding SWIT in \autoref{fig:combined-dag}(ii), we see that $A_1^{a_0}$ is not d-separated from $A_0$ given $Y_0$. Since joint independence implies marginal independence, it follows that $\left(Y_1^{a_0}, A_1^{a_0}, Y_2^{\bar{a}_1}\right)$ is not independent from $A_0$ given $Y_0$, and therefore, exchangeability (\ref{ass:ex:formula}) does not hold.

\newpage
\noindent Finally, we impose the following positivity assumption. 
\begin{assumption}[Positivity]
\begin{align}
\label{ass:pos}
\begin{split}
    & f_{\bar{\boldsymbol{L}}_k, \bar{Y}_k, \bar{A}_{k}}(\bar{\boldsymbol{l}}_k, \bar{y}_k, \bar{a}_k) > 0  \; \Rightarrow \; f_{A_k \mid  \bar{\boldsymbol{L}}_k, \bar{Y}_k, \bar{A}_{k-1}}(g_{j}(y_k, a_k) \mid\bar{\boldsymbol{l}}_k, \bar{y}_k, \bar{a}_{k-1}) > 0,
\end{split}
\end{align}
for all $j \in \{0,1\}$, $\bar{\boldsymbol{l}}_k \in \mathcal{L}^{k+1}$, $\bar{y}_k \in \mathcal{Y}^{k+1}$, $\bar{a}_k \in \{0,1\}^{k+1}$, and $k \leq \kappa$. 
\end{assumption}
\noindent Positivity requires that, at each time point $k$, and for every possible history of covariates and treatments up to $k$, there is a positive probability of following both the add-on-1 and the add-on-0 regime. This assumption holds by design in randomized trials, where each participant has a known probability of following either regime. In observational studies, violations of positivity can be structural or random. Structural violations occur when individuals at certain levels of a confounder cannot possibly follow a regime. For example, heart conditions may influence opioid use, and conditions like heart failure contraindicate NSAIDs. If the probability of receiving NSAIDs among individuals with heart failure is zero, then any person with heart failure who receives opioids cannot follow the add-on-1 regime, resulting in a structural violation of positivity \parencite{westreich2010invited}. Random violations arise when, purely by chance, no individuals at certain confounder levels follow one of the regimes.

\subsection{Identification Results}

\noindent If consistency (\ref{ass:con}), positivity (\ref{ass:pos}), and exchangeability (\ref{ass:ex:formula}) hold, the expected opioid dose under the add-on regime can be identified using the following g-formula \parencite{robins1986new,richardson2013single}.

\begin{theorem}[G-formula]
\label{g:theorem}
Let $g_j$ be an add-on-j regime and assume that consistency (\ref{ass:con}), positivity (\ref{ass:pos}), and exchangeability (\ref{ass:ex:formula}) hold. Then
   \begin{align}
   \label{g:formula}
   \begin{split}
     E\left[ Y_k^{g_j} \right]  = & \int \ldots \int E[Y_k \mid\bar{\boldsymbol{L}}_{k'} = \bar{\boldsymbol{l}}_{k'}, \bar{Y}_{k'} = \bar{y}_{k'}, A_0 = g_j(y_0, a_0), \ldots, A_{k'} = g_j(y_{k'}, a_{k'})] \\
    & \prod_{t = 1}^{k'} f_{\boldsymbol{L}_{t}, Y_{t}, A_{t} \mid\bar{\boldsymbol{L}}_{t-1}, \bar{Y}_{t-1}, \bar{A}_{t-1}}(\boldsymbol{l}_{t}, y_{t}, a_{t} \mid\bar{\boldsymbol{l}}_{t-1}, \bar{y}_{t-1}, g_j(y_0, a_0),  \ldots, g_j(y_{t-1}, a_{t-1})) \\ 
    & f_{\boldsymbol{L}_0, Y_0, A_0}(\boldsymbol{l}_0, y_0, a_0) \, \textup{d} \, \boldsymbol{l}_{k'} \, \textup{d} \, y_{k'} \, \textup{d} \, a_{k'} \ldots \textup{d} \, \boldsymbol{l}_0 \, \textup{d} \, y_0 \, \textup{d} \, a_0, 
   \end{split}
   \end{align}
for all $j \in \{0,1\}$ and $k \geq 1$ where $k' = \min \{k-1, \kappa\}.$
\end{theorem}

\noindent In Appendix A, we demonstrate that the identification result in \autoref{g:formula} is a special case of the g-formula presented in Equation ($67$) of \textcite[p.~67]{richardson2013single}. \\
\indent As an alternative identification result, we provide an inverse probability weighted (IPW) identification formula \parencite{robins1992recovery, rotnitzky1995semiparametric}. 

\begin{corollary}[IPW] 
\label{ipw:corollary}
Let $g_j$ be an add-on-$j$ regime and assume that consistency (\ref{ass:con}), positivity (\ref{ass:pos}), and exchangeability (\ref{ass:ex:formula}) hold. Then
   \begin{align}
   \label{ipw:formula}
   \begin{split}
   E\left[ Y_k^{g_j} \right] = E[Y_k \, W_{ k' }], 
   \end{split}
   \end{align}
for all $j \in \{0,1\}$ and $k \geq 1$ where $k' = \min \{k-1, \kappa\}$ and 
\begin{align}
\label{ipw:weights}
        W_{s} = \sum_{a_0 = 0}^1 \ldots \sum_{a_{s} = 0}^1 \prod_{t = 0}^{s}\frac{ 1\{A_t = g_j(Y_t, a_t)\} f_{A_t|\bar{\boldsymbol{L}}_t, \bar{Y}_t, \bar{A}_{t-1}}(a_t|\bar{\boldsymbol{L}}_t, \bar{Y}_t, \bar{A}_{t-1})}{f_{A_t \mid\bar{\boldsymbol{L}}_t, \bar{Y}_t, \bar{A}_{t-1}}(A_t \mid\bar{\boldsymbol{L}}_t, \bar{Y}_t, \bar{A}_{t-1})} \quad \forall \, s \geq 0. 
\end{align}
\end{corollary}

\noindent In Appendix A, we demonstrate that the inverse probability weighted identification formula in \autoref{ipw:formula} is a special case of the inverse probability weighted identification formula in Equation $(68)$ of \textcite[p.~67]{richardson2013single}. 

\section{Censoring and Competing Events}
\label{sec:censor}

Here we extend the results to account for censoring and competing events; additional elaborations and proofs are given in Appendix B. A competing event is an event that precludes the occurrence of the event of interest. In our application, any event that deterministically sets all future opioid doses to zero is considered a competing event with respect to the event of interest, opioid dose. For example, death is a competing event with respect to opioid dose, as an individual cannot use opioids after death. For clarity, we formalize this in the following definition.

\begin{definition}\label{def:competing} Let $(D_{k})_{k \leq K}$ be a sequence of random variables where $P(D_0 = 0) = 1$,  $P(D_k \in \{0,1\}) = 1$ for every $k \leq K$, and $D_k = 1$ imples $\underline{D}_k = \bar{1}$ for every $k \leq K$. If $D_k = 1$ implies $\underline{Y}_k = \bar{0}$ for every $k \leq K$, we say that $(D_{k})_{k \leq K}$ is a sequence of competing event indicators with respect to $(Y_k)_{k \leq K}$ with $D_k$ indicating competing event by time $k$. 
\end{definition}
\noindent Consider the variables 
\begin{align}
\label{framework:variables:latent:cd}
    (D_0, \boldsymbol{L}_0, Y_0, A_0, \ldots, D_K, \boldsymbol{L}_K, Y_K, A_K),
\end{align}
where $\boldsymbol{L}_k, Y_k$, and $A_k$ are defined as in \autoref{sec:observeddata} for every $k \leq K$ and $(D_k)_{k \leq K}$ is a sequence of competing event indicators with respect to $(Y_k)_{k \leq K}$. The variables are topologically ordered according to (\ref{framework:variables:latent:cd}). \\
\indent When censoring exists, the variables in (\ref{framework:variables:latent:cd}) are not fully observable and are therefore referred to as latent variables. We use a tilde to distinguish the potentially censored variables observed in the data from their latent counterparts in (\ref{framework:variables:latent:cd}). Specifically, consider the variables $(\tilde{D}_0, \tilde{L}_0, \tilde{Y}_0, \tilde{A}_0, \ldots, \tilde{D}_K, \tilde{L}_K, \tilde{Y}_K, \tilde{A}_K)$ where $\tilde{D}_k \in \{0,1\}$ is a potentially censored indicator of competing event by time $k$, $\tilde{L}_k \in \mathcal{L}$ is a vector of potentially censored covariates at time $k$, $\tilde{Y}_k \in \mathcal{Y}$ denotes the potentially censored opioid dose at time $k$, and $\tilde{A}_k \in \{0,1\}$ indicates the potentially censored NSAID treatment at time $k$ for every $k \leq K.$ \\
\indent We define a censoring event as an event whose absence ensures that the potentially censored variables align with their latent counterparts. For instance, loss to follow-up is commonly treated as a censoring event because when it does not occur, the variables recorded in the data are assumed to coincide with their latent counterparts. For clarity, we formalize this in the following definition.

\begin{definition}\label{def:censoring} Let $(C_{k})_{k \leq K}$ be a sequence of random variables where $P(C_0 = 0) = 1$, $P(C_k \in \{0,1\}) = 1$ for every $k \leq K$ and $C_k = 1$ implies $\underline{C}_k = \bar{1}$ for every $k \leq K$. If $C_k = 0$ implies $\tilde{D}_k = D_k, \tilde{L}_k = \boldsymbol{L}_k, \tilde{Y}_k = Y_k$, and $\tilde{A}_k = A_k$ for every $k \leq K$, we say that $(C_{k})_{k \leq K}$ is a sequence of censoring indicators with $C_k$ indicating censoring by time $k$. 
\end{definition}

\noindent In the following, let $(C_{k})_{k \leq K}$ be a sequence of censoring indicators and assume that we have access to $n$ realizations of independent replicates of 
\begin{align}
\label{framework:variables:cd}
    (C_0, \tilde{D}_0, \tilde{L}_0, \tilde{Y}_0, \tilde{A}_0, \ldots, C_K, \tilde{D}_K, \tilde{L}_K, \tilde{Y}_K, \tilde{A}_K),
\end{align}
corresponding to study patients randomly sampled from a target population of interest. The variables are topologically ordered according to $(\ref{framework:variables:cd})$. We denote counterfactual variables by superscripts. For example, $Y_k^{g, \bar{c} =  0}$ represents the opioid dose that would have been observed at time $k$ under a given treatment regime $g$, in the absence of censoring ($\bar{C}_k = \bar{0}$) by time $k$. A formal presentation of all counterfactual variables of interest is provided in Appendix B. \\
\indent When censoring and competing events are present, the counterfactual outcomes—and therefore the causal estimands—can be defined in multiple ways. See \textcite{young2020causal} for a thorough presentation. Here, we define the add-on effect as the average causal effect of the add-on-1 regime $g_1$ versus the add-on-0 regime $g_0$, under an additional intervention that eliminates censoring, on opioid dose through all causal pathways, including pathways through the competing event. Therefore, a non-null value of the opioid-sparing effect is not sufficient to conclude that the add-on regime exerts a direct effect on the total opioid dose outside of the competing event. The opioid-sparing effect may also, or exclusively, reflect an effect on the competing event, influencing the duration of time individuals remain at risk of opioid use.

\begin{definition}[Add-on effect] 
\begin{align}
\label{eff:ose:cd}
         E\left[ Y_k^{g_1, \bar{c} = \bar{0}} \right] - E\left[ Y_k^{g_0, \bar{c} = \bar{0}} \right] \quad \forall \, k \geq 1.
\end{align}
\end{definition}

\noindent The identification of the add-on effect (\ref{eff:ose:cd}) in the presence of censoring and competing events proceeds analogously to the identification of the add-on effect (\ref{eff:aose:timek}) in the absence of censoring and competing events, with the main difference being that we also intervene to eliminate censoring and include the competing event as a component of the covariates. Proofs are provided in Appendix B.

\section{Estimation}
\label{sec:estimation}

\noindent The g-formula, given by the right-hand side of (\ref{g:formula}), can be estimated using standard models and software, e.g., the \code{gfoRmula} package in \code{R} \parencite{mcgrath2020gformula}. This package estimates the required conditional outcome means and covariate distributions using parametric models, and approximates the resulting expectations via Monte Carlo simulation. Further details are provided in \textcite{mcgrath2020gformula}. An application is presented in \autoref{sec:application}, with corresponding code included in the \href{https://github.com/CatharinaStoltenberg/add-on-effects}{supplementary material}. \\
\indent The inverse probability weighted identification formula, given by the right-hand side of (\ref{ipw:formula}), can be estimated in two steps using standard \code{R} functions. First, estimate the conditional probability of treatment given the observed covariate and treatment history to obtain the inverse probability weights defined in (\ref{ipw:weights}). This can be implemented by fitting a logistic regression model for the probability of treatment given the observed history, using iterative weighted least squares via the \code{glm} function in \code{R}. Second, fit a weighted model for the outcome, using the weights obtained in the first step. This can be implemented by fitting a linear model for the outcome by weighted least squares, again using \code{glm}, with individuals weighted by their estimated inverse probability weights. An application is presented in \autoref{sec:application}, with corresponding code included in the \href{https://github.com/CatharinaStoltenberg/add-on-effects}{supplementary material}. \\
\indent It is also possible to construct doubly robust estimators \parencite{robins1994estimation, robins2000robust, laan2003unified, bang2005doubly}. That is, estimators that are consistent if, at each time point, either the outcome mechanism or the treatment mechanism is consistently estimated. In Appendix C, we show that the g-formula in \autoref{g:formula} is a special case of the identification formula stated in Theorem 1 of \textcite{diaz2023nonparametric}, but obtained under different identifiability assumptions. However, this correspondence enables the application of Proposition 1 in \textcite{diaz2023nonparametric} to construct sequentially doubly robust estimators, such as the one described in Section 5.3 of their work. These estimators can account for both censoring and competing events and can be implemented using the R package \code{lmtp} \parencite{williams2023lmtp}.

\section{Application: Opioid-Sparing Effects of NSAIDs in Trauma Patients Initially Treated with Opioids}  
\label{sec:application} 

\begin{figure}[h!]
   \centering
    \includegraphics[width=200em, height = 25em, keepaspectratio]{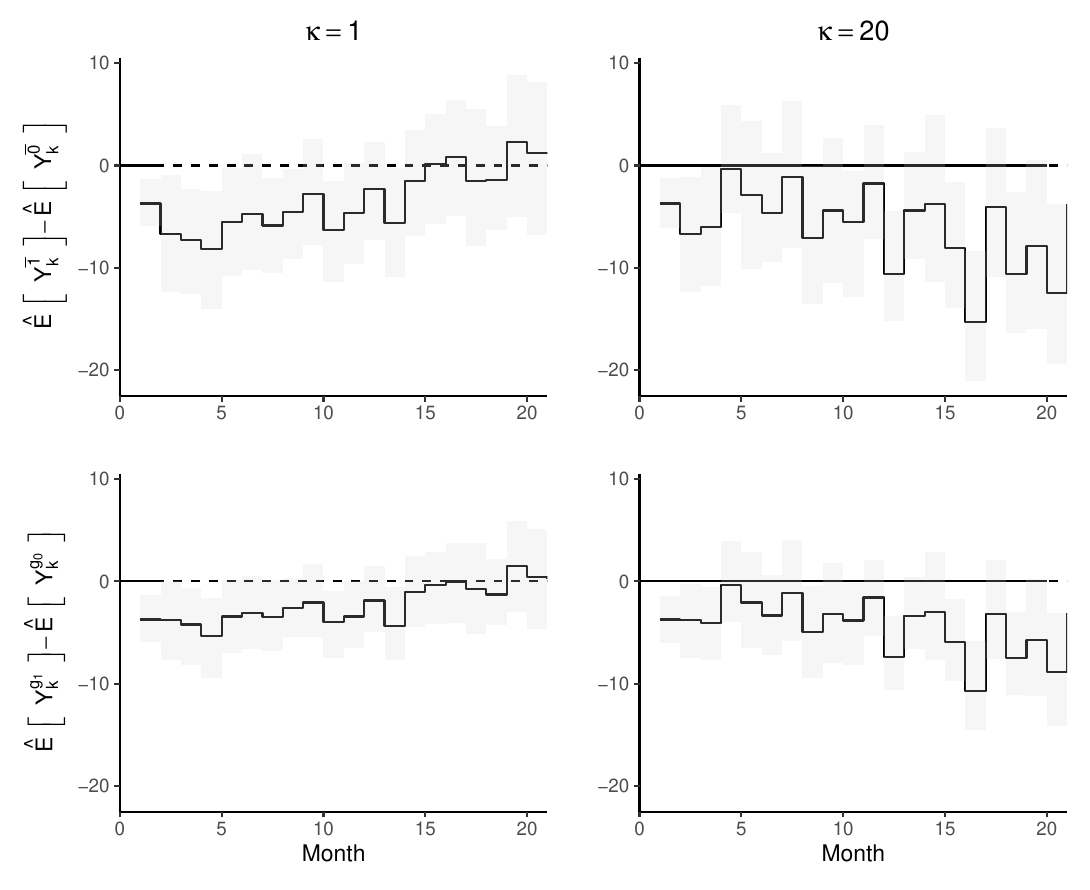}
    \caption{G-formula estimates of the expected difference in monthly dispensed opioid dose under four treatment comparisons: static regime $(a_0, a_1) = (1,1)$ vs.\ $(a_0, a_1) = (0,0)$ (top left) and add-on-$1$ regime $g_1$ vs.\ add-on-$0$ regime $g_0$ (bottom left), both with a short treatment period with $\kappa = 1$; static regime $(a_0, \ldots, a_{20}) = (1, \ldots, 1)$ vs.\ $(a_0, \ldots, a_{20}) =(0, \ldots, 0)$ (top right) and add-on-$1$ vs.\ add-on-$0$ (bottom right), both with a long treatment period with $\kappa = 20$. Percentile $95\%$ confidence intervals were obtained via nonparametric bootstrap with $500$ samples.}
\label{app:figure1}
\end{figure}

\noindent We estimated the opioid-sparing effect of NSAIDs in a cohort of trauma patients initially treated with opioids. The analysis used observational data from \href{https://www.ous-research.no/home/ipot/Projects/20448}{NTRplus}, which links the \href{https://www.ous-research.no/home/ipot/Projects/20448}{Norwegian National Trauma Registry} to several national databases: the \href{https://www.norpd.no/}{Norwegian Prescription Database}, the \href{https://www.fhi.no/en/ch/cause-of-death-registry/}{Cause of Death Registry}, the \href{https://helsedata.no/en/forvaltere/norwegian-institute-of-public-health/norwegian-patient-registry-npr/}{Norwegian Patient Registry}, and \href{https://www.ssb.no/en}{Statistics Norway} \parencite{ringdal2025norwegian}. Dispensation records for NSAIDs and opioids were obtained from the Norwegian Prescription Database, which includes all prescription-based dispensations. Over-the-counter and dispensations from institutions such as hospitals or nursing homes were not included. Opioid dose was measured in Oral Morphine Equivalents (OMEQ).\\
\indent We defined follow-up time in months $k \in \{0, \ldots, 21\}$, where month $0$ corresponds to the first month after the initial opioid dispensation within the first post-discharge month, and month $21$ marks the end of follow-up. We considered two treatment periods: on short $\{0,1\}$ with $\kappa = 1$ and one long $\{0,1,\ldots, 20\}$ with $\kappa = 20$. No individuals were censored during follow-up.\\
\indent The study cohort included $8,\!718$ individuals with potentially serious injuries treated by the specialist health services in Norway, including prehospital care, from 2015 to 2018. Inclusion criteria were: (i) registered discharge date, (ii) eligibility for NSAIDs, (iii) at least one opioid dispensation during the first month post-discharge following the first registered traumatic injury, and (iv) survival through month $0$. The data span the years 2014–2020.  \\
\indent Under the add-on-$j$ regime, if opioids are dispensed by prescription in a given month of the treatment period, NSAIDs are also dispensed for $j = 1$ and not dispensed for $j = 0$. If no opioids are dispensed, NSAIDs are dispensed according to their natural level that month.\\
\indent The analysis adjusted for both baseline and time-varying covariates. Baseline covariates included demographic factors (age, sex, income, and residential region), pre-injury medication administration (NSAID dispensations and opioid dispensations within six months prior to month $0$), injury-related factors (mechanism of injury, type of accident, American Society of Anesthesiologists physical status classification, and the Glasgow Outcome Scale at discharge), and hospital-related characteristics (highest level of care received, admission to an intensive care unit, and total length of hospital stay). Time-varying covariates included healthcare utilization, concurrent use of medications, and opioid dispensations over time. A complete description of the variables, the target trial, and its observational emulation is provided in Appendix D. \\
\indent To estimate the add-on effect of NSAIDs on monthly opioid dose (\ref{eff:aose:timek}) and on the total opioid dose (\ref{eff:aose}), we implemented the g-formula estimator corresponding to the right-hand side of (\ref{g:formula}), using the \code{gfoRmula} package in \code{R} \parencite{mcgrath2020gformula}. We also estimated the expected difference in monthly and total opioid dose between the static regimes $\bar{a}_{\kappa} = \bar{1}$ (NSAIDs dispensed at every time-point during the treatment period) and $\bar{a}_{\kappa} = \bar{0}$ (NSAIDs never dispensed during the treatment period). The full implementation code is provided in the \href{https://github.com/CatharinaStoltenberg/add-on-effects}{supplementary material}.\\
\indent All point estimates for the effect on total opioid dose were negative, indicating an opioid-sparing effect of NSAIDs (See \autoref{tab:simple}). For the long treatment period with $\kappa = 20$, the $95\%$-confidence intervals excluded zero; for the short treatment period with $\kappa = 1$, the $95\%$-confidence intervals included zero. \\
\indent For the effect on monthly opioid dose, three patterns emerged (See \autoref{app:figure1}). First, for the short treatment period with $\kappa = 1$, the effect was negative early in follow-up but gradually approached zero, suggesting a short-term opioid-sparing effect that diminished over time. Second, for the long treatment period with $\kappa = 20$, the effect remained negative throughout follow-up, indicating a more sustained effect. Third, the effect was more pronounced under static regimes than add-on regimes. Still, the add-on regime remains more clinically relevant, as the static regime enforces NSAID use regardless of clinical need. Overall, the results support an opioid-sparing effect of NSAIDs.

\vspace{2 em}

\begin{table}[h!]
\caption{G-formula estimates of the expected difference in total monthly dispensed opioid dose under different treatment regimes and treatment periods. Percentile $95\%$ confidence intervals were obtained via nonparametric bootstrap with $500$ samples.}  

\label{app:table}
    \centering
    \begin{tabular}{lll}
        \toprule
        \textbf{Parameter} &  \textbf{Treatment period} & \textbf{Estimate ($95\%$ CI)} \\
        \midrule
        $\sum_{k=1}^{21} \left( E\left[ Y_k^{\bar{a} = \bar{1}} \right] - E\left[ Y_k^{\bar{a} = \bar{0}} \right] \right)$ & $\{0,1\}$ & $-66.96$ $(-161.26, 27.89)$ \\
        $\sum_{k=1}^{21} \left( E\left[ Y_k^{g_1} \right] - E\left[ Y_k^{g_0} \right] \right)$  & $\{0,1\}$ & $-46.77$ $(-104.23, 13.33)$ \\
        {\scriptsize$\sum_{k=1}^{21} \left( E\left[ Y_k^{\bar{a} = \bar{1}} \right] - E\left[ Y_k^{\bar{a} = \bar{0}} \right] \right)$} & $\{0, \ldots, 20\}$ & $-125.16$ $(-173.34, -64.63)$ \\
        $\sum_{k=1}^{21} \left( E\left[ Y_k^{g_1} \right] - E\left[ Y_k^{g_0} \right] \right)$  & $\{0, \ldots, 20\}$ & $-90.99$ $(-126.42, -49.89)$ \\
        \bottomrule
    \end{tabular}
    \label{tab:simple}
\end{table} 

\vspace{2 em}

\section{Discussion}

The add-on regime enables us to define the effect of adding one intervention whenever another is naturally implemented on an outcome of interest. In our motivating example on opioid-sparing treatments, it enables us to define the opioid-sparing effect of supplementing opioid treatments with additional analgesics, such as NSAIDs, on total opioid consumption. Quantifying this effect is clinically important, as it can inform strategies to reduce opioid reliance in pain management. \\
\indent We derived identification results consistent with the more general results of \textcite[p.~67]{richardson2013single}, but under a distinct set of conditions. In particular, we argue that the assumptions used to prove the identification results presented here are easier to assess in practice than those used in \textcite[p.~67]{richardson2013single}. \\
\indent Another natural extension of the work presented in this paper is to allow interventions on both treatment components and generalize to non-binary treatments. In the context of opioid-sparing strategies, this would enable the evaluation of regimes in which a specified dose of NSAIDs, potentially defined relatively to the natural opioid dose, is added whenever opioids are naturally administered, and the opioid dose is reduced by a fixed proportion of its natural value. 

\section{Supplementary Materials}

The \href{https://github.com/CatharinaStoltenberg/add-on-effects}{supplementary materials} include an appendix containing formal definitions of counterfactual variables and proofs of the identification results. They also demonstrate how these findings relate to established, more general results, including a proof that the exchangeability condition used in those results implies the version applied in this manuscript. Additionally, the appendix provides detailed information on all variables used in the analysis, as well as a summary of the target trial protocol and its corresponding observational emulation. The supplementary materials also include the \code{R} computer code used for the analysis.

\section{Acknowledgement}

We thank Statistics Norway, the National Institute of Health, and the Directorate of Health for providing access to data. This publication uses data from the Norwegian Patient Registry and several other national registries. The interpretation and reporting of these data are the sole responsibility of the authors and do not imply endorsement by the registries. We are also grateful to the trauma hospital registrars in Norway for their meticulous documentation, which ensures the high quality of the Norwegian Trauma Registry. Finally, we thank the patients included in the registry.

\section{Data Availability Statement}

The analysis used observational data from the \href{https://www.ous-research.no/home/ipot/Projects/20448}{Norwegian National Trauma Registry}, the \href{https://www.norpd.no/}{Norwegian Prescription Database}, the \href{https://www.fhi.no/en/ch/cause-of-death-registry/}{Cause of Death Registry}, the \href{https://helsedata.no/en/forvaltere/norwegian-institute-of-public-health/norwegian-patient-registry-npr/}{Norwegian Patient Registry}, and \href{https://www.ssb.no/en}{Statistics Norway} \parencite{ringdal2025norwegian}. This data is not publicly available due to privacy regulations. Access to the data can be applied for through the respective data holders, subject to ethical approval and data protection requirements.

\section{Disclosure Statement}

The authors report that there are no competing interests to declare. This manuscript includes minor language corrections assisted by ChatGPT (OpenAI, GPT-4, July 2025 version). The tool was used to improve clarity and grammar during the revision process. All intellectual content, interpretation, and conclusions are the author’s own.

\section{Funding}

This work was conducted while Catharina Stoltenberg was a Ph.D. student at the Oslo Center for Biostatistics and Epidemiology, University of Oslo. It was supported by the Institute of Basic Medical Sciences at the University of Oslo, the Research Council of Norway, and the Department of Research and Development, Division of Emergencies and Critical Care, Oslo University Hospital. Mats Stensrud was supported by the Swiss National Science Foundation. 


\newpage

\printbibliography

\end{document}